\documentclass[12pt,preprint]{aastex}

\usepackage{amsmath}
\usepackage{amssymb}
\usepackage{bm}

\shorttitle{The power-law spectra during magnetic reconnection}
\shortauthors{Drake et al}

\begin{document}


\title{The power-law spectra of energetic particles during multi-island magnetic reconnection}


\author{
J.~F.~Drake\altaffilmark{1}, %\\
M.~Swisdak\altaffilmark{1}, %\\
R.~Fermo\altaffilmark{2}
}


\altaffiltext{1}{University of Maryland, College Park, MD, USA;
 drake@umd.edu, swisdak@umd.edu}
\altaffiltext{2}{Astronomy Department, Boston
University, Boston, MA, USA ; rfermo@bu.edu}


\begin{abstract}

Power-law distributions are a near universal feature of energetic
particle spectra in the heliosphere. Anomalous Cosmic Rays (ACRs),
super-Alfv\'enic ions in the solar wind and the hardest energetic
electron spectra in flares all have energy fluxes with power-laws that
depend on energy $E$ approximately as $E^{-1.5}$. We present a new
model of particle acceleration in systems with a bath of merging
magnetic islands that self-consistently describes the development of
velocity-space anisotropy parallel and perpendicular to the local
magnetic field and includes the self-consistent feedback of pressure
anisotropy on the merging dynamics. By including pitch-angle
scattering we obtain an equation for the omni-directional particle
distribution $f(v,t)$ that is solved in closed form to reveal $v^{-5}$
(corresponding to an energy flux varying as $E^{-1.5}$) as a
near-universal solution as long as the characteristic acceleration time is
short compared with the characteristic loss time. In such a state the
total energy in the energetic particles reaches parity with the
remaining magnetic free energy. More generally, the resulting
transport equation can serve as the basis for calculating the
distribution of energetic particles resulting from reconnection in
large-scale inhomogeneous systems.

\end{abstract}




\section{INTRODUCTION}\label{intro}
Accelerated particles with power-law spectra are a nearly universal
feature of heliospheric plasmas and also characterize the cosmic ray
spectrum. Anomalous Cosmic Rays (ACRs) \citep{Stone08,Decker10},
super-Alfv\'enic ions in the solar wind \citep{Fisk06} and the hardest
energetic electron spectra in flares \citep{Holman03} all have energy
fluxes with power-laws that depend on energy $E$ approximately as
$E^{-1.5}$. An important question is whether there is a common
acceleration mechanism in these very disparate environments.

A range of acceleration mechanisms have been proposed to explain the
spectra of energetic electrons (up to several $MeV$) and ions (up to
several $GeV$) in impulsive flares, including the reconnection process
itself and reconnection-driven turbulence
\citep{Miller97,Dmitruk04,Liu06,Zharkova11}. Significant challenges
have been to explain the large numbers of accelerated electrons and
the surprising efficiency of the conversion of magnetic energy to the
energetic particles \citep{Lin71,Emslie05,Krucker10}. The single
x-line model of reconnection in flares, in which electrons are
accelerated by parallel electric fields, can not explain the large
number of accelerated electrons \citep{Miller97}. On the other hand
both observations \citep{Sheeley04,Savage12} and modeling
\citep{Kliem94,Shibata01,Drake06,Drake06a,Onofri06,Oka10,Huang11,Daughton11,Fermo12}
suggest that reconnection in flares involves the dynamics of large
numbers of x-lines and magnetic islands or flux ropes. In
magnetohydrodynamic (MHD) simulations of multi-island reconnection
test particles rapidly gain more energy than is available in the
driving magnetic field \citep{Onofri06}. Thus, developing a model of
particle acceleration in a multi-island reconnecting environment with
feedback on the driving fields is the key to understanding
flare-produced energetic particle spectra.

The seed population of ACRs are interstellar pickup particles since
their composition matches that of interstellar neutrals
\citep{Cummings96,Cummings07}. However, the conventional idea that
they are accelerated at the termination shock (TS)\citep{Pesses81} was
called into question when the Voyagers crossed the TS and found that
the intensity of the ACR spectrum did not peak there
\citep{Stone05,Stone08}. A possible alternate source is magnetic reconnection
of the sectored heliosheath \citep{Lazarian09,Drake10}. Simulations of
reconnection in the sectored field region revealed that
the dominant heating mechanism was Fermi reflection in contracting and
merging islands \citep{Drake10,Kowal11,Schoeffler11}. Because contraction
increases the energy parallel to the local magnetic field and
reduces the perpendicular energy, the heating mechanism drives the system to
the firehose stability boundary 
$\alpha=1-(\beta_\parallel -\beta_\perp) /2=0$ where
reconnection is throttled because the magnetic
tension drive is absent \citep{Drake06,Drake10,Opher11,Schoeffler11}.

While multi-island simulations have revealed that Fermi reflection in
contracting islands controls energy gain and drives the system to the
marginal firehose condition, a rigorous model for particle
acceleration in such a multi-island system has not yet been
developed. The Parker equation does not describe particle acceleration
in nearly incompressible systems \citep{Parker65} and extensions do
not account for the geometry of reconnection and island merging
\citep{Earl88}. In the present manuscript we explore
particle acceleration in a bath of merging magnetic islands with a
particle distribution function $f(v_\parallel,v_\perp)$ that accounts for
the velocity space anisotropy along ($v_\parallel$) and across
($v_\perp$) the local magnetic field and includes a phenomenological
pitch-angle scattering operator. Thus, the pressure anisotropy can be
directly evaluated and the feedback on island merging calculated. 

\section{Particle dynamics during island merger}

\begin{figure}
\epsscale{.70}
\includegraphics[keepaspectratio,width=4.0in]{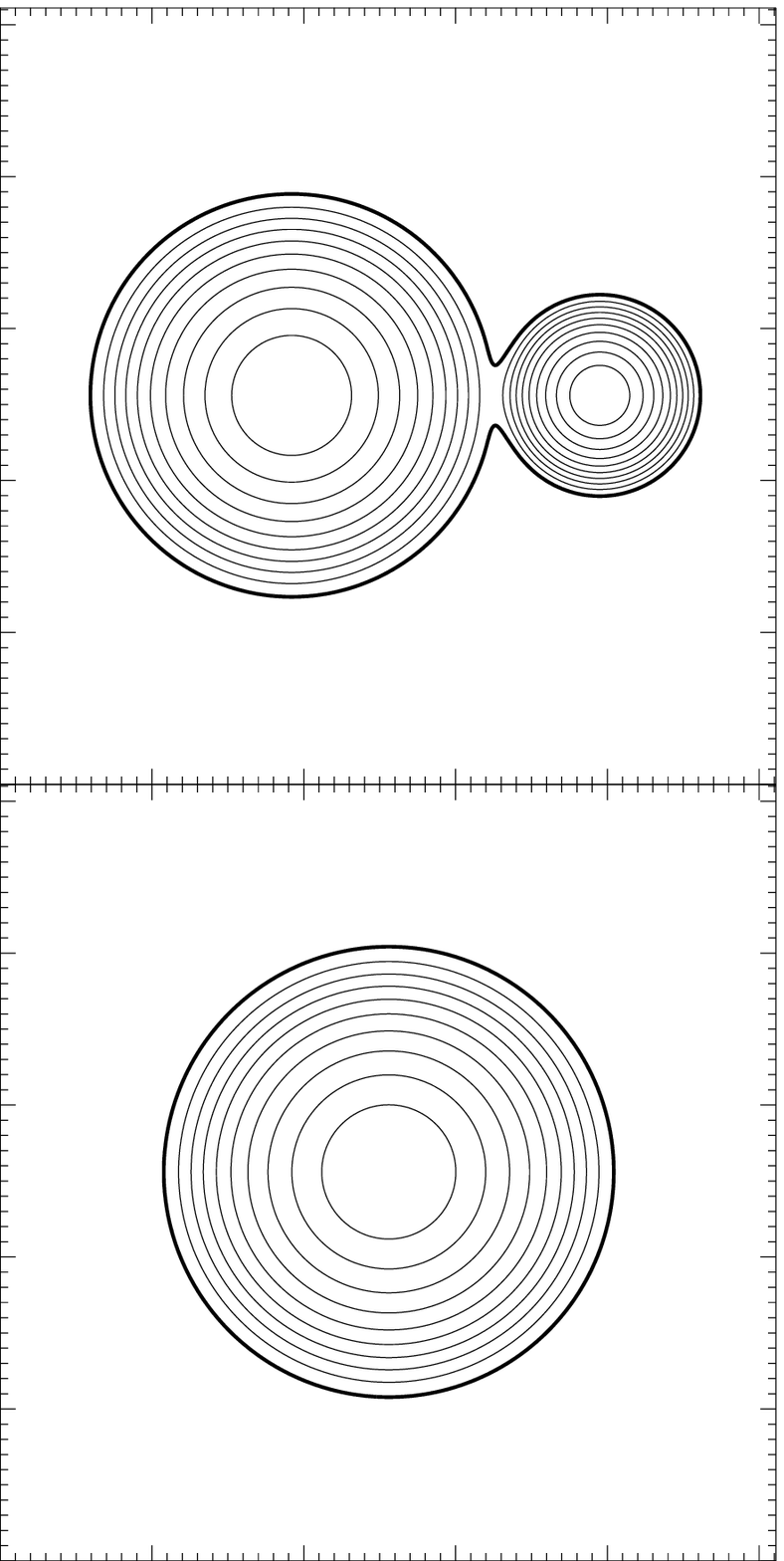}
\caption{\label{islandmerger} Schematic of the merger of two islands
  of differing radii and magnetic field strength.}
%\plotone{JandVX.pdf}
%\includegraphics[height=220mm,width=180mm]{JandVX.pdf}
\end{figure}

We develop a probabilistic model of particle acceleration in a bath of
merging 2-D magnetic islands with a distribution of magnetic flux
$\psi$ and area $A$ given by $g(\psi , A)$ \citep{Fermo10}. The
development of structure in 3-D may ultimately be important and should
be addressed but observations \citep{Phan06} and simulations
\citep{Hesse01} suggest that at the largest scales reconnection is
nearly 2-D and this limit is therefore a reasonable starting point. We
first calculate the particle energy gain during the merging of two
circular islands of radii $r_1$ and $r_2$ with $r_j=\sqrt{A_j/\pi}$ as
shown in the schematic in Fig.~\ref{islandmerger}. Merging leads to a
single island of area $A_f=A_1+A_2$ and with magnetic flux $\psi_f$
given by the larger of $\psi_1$ and $\psi_2$ \citep{Fermo10}. The
reduction of energy by the factor $(\psi_1^2+\psi_2^2)/\psi_f^2$
results from the shortening of the field lines as merging
proceeds. Thus, energy release takes place not at the merging site,
but as reconnected field lines contract after merger. As long as the
kinetic-scale, boundary layer where reconnection occurs is small
compared with the island radii, the dominant energy exchange with
particles takes place on the closed, reconnected field lines that
release magnetic energy as they contract.

We take advantage of two adiabatic invariants, the magnetic moment
$\mu=mv_\perp^2/B$ and the parallel action $\oint v_\parallel d\ell$,
which are constants if the gyration time of particles around the local
magnetic field and their circulation time around islands are short
compared with the merging time. The former describes the reduction in
$v_\perp$ as $B$ decreases and the latter the increase in
$v_\parallel$ as $\ell$ decreases. The parallel action invariant is
valid for velocities that exceed the local Alfv\'en speed, which
implies that a seed heating mechanism is needed for low $\beta$ systems
such as the solar corona \citep{Drake09a,Knizhnik11} but not in high
$\beta$ systems such as the sectored heliosheath.  To calculate
$\dot{\ell}$, we first calculate the merging velocity $\dot{r}_{sep}$
of two islands with differing radii and magnetic fields,
$\dot{r}_{sep}=\dot{r}_1+\dot{r}_2=-\dot{\psi}(B_1+B_2)/(B_1B_2)$,
since merging magnetic islands reconnect their magnetic flux at the
same rate. The reconnection rate is given by \cite{Cassak07c},
$\dot{\psi}=2V_{12}B_1B_2/(B_1+B_2)$, with
$V_{12}=\epsilon_r\sqrt{\alpha_{12}B_1B_2/4\pi\rho}$, where
$\epsilon_r\sim 0.1$ is the normalized rate of reconnection and
$\alpha_{12}=1-4\pi (p_\parallel -p_\perp )/(B_1B_2)$. Thus,
$\dot{r}_{sep}=-2V_{12}$ and $V_{12}$ is the island merging
velocity. The rate of line shortening can now be calculated from the
total merging time $(r_1+r_2)/(2V_{12})$ and the difference between
the intial field line length as merging starts and the final length
using area conservation, $\dot{\ell}=-2\pi h_{12}V_{12}$ with
$h_{12}=2(r_1+r_2-\sqrt{r_1^2+r_2^2})/(r_1+r_2)$. Parallel action
conservation then yields an equation for $v_\parallel$,
\begin{equation}
\frac{dv_\parallel}{dt}=v_\parallel\frac{h_{12}V_{12}}{r_1+r_2}.
\label{vpardot}
\end{equation}
To obtain the corresponding equation for $\dot{v}_\perp$, we use the conservation of magnetic flux and area as a flux tube contracts so that $B/\ell$ is constant. Therefore, from $\mu$ conservation $v_\perp^2/\ell$ is also constant and 
\begin{equation}
\frac{dv_\perp^2}{dt}=-v_\perp^2\frac{h_{12}V_{12}}{r_1+r_2}.
\label{vperpdot}
\end{equation}
Thus, the perpendicular energy goes down during island merger as the parallel energy increases. 

\section{A kinetic equation for particle acceleration during island merger}
From the energy gain of particles in merging islands we can formulate
a model of particle acceleration in a very long current
layer of length $L$. Particles are injected into the bath of
interacting islands in the current layer from upstream as each
individual island grows due to reconnection of the upstream field. They
then undergo acceleration in the merging islands until they are
convectively lost. The rate of injection of particles is given by the
upstream distribution function $f_{up}(v)$ times the integrated rate
of area increase of all of the magnetic islands $\dot{A}_T$
\citep{Fermo10},
\begin{equation}
\dot{A}_T=2\pi\epsilon_rc_{Aup}\int_0^\infty \int_0^\infty dAd\psi \, rg(\psi , A),
\label{Adot}
\end{equation} 
with the island radius given by $r=\sqrt{A/\pi}$. The probability of two islands of radii $r_1$ and $r_2$ merging is given by their overlap probability $4r_1r_2/L^2$. Using the conservation of phase space volume and summing over all merging islands in the layer, we obtain a differential equation for $f(v_\parallel , v_\perp)$,
\begin{equation}
\frac{\partial f}{\partial t}+R\left ( \frac{\partial}{\partial
  v_\parallel}v_\parallel-\frac{1}{2v_\perp}\frac{\partial}{\partial
  v_\perp}v_\perp^2\right ) f
-\nu\frac{\partial}{\partial\zeta}\left (1-\zeta^2\right )\frac{\partial}{\partial\zeta}f+\frac{c_{Aup}}{L}f=\dot{A}_Tf_{up},
\label{fequation}
\end{equation}
where 
\begin{equation}
R=\int\int d1d2\,g_1g_2\frac{4r_1r_2h_{12}V_{12}}{L^2(r_1+r_2)}.
\label{rdrive}
\end{equation}
with $di=dA_id\psi_i$. In earlier simulations of multi-island
reconnection strong pressure anisotropy with $p_\parallel > p_\perp$
within the core of merging islands was limited by anisotropy
instabilities \citep{Drake10,Schoeffler11} so we have included a
phenomenological pitch-angle scattering operator of strength $\nu$
that acts on the angle $\zeta = v_\parallel /v$ to reduce
anisotropy. Importantly, the drive $R$ is independent of the particle
velocity. It depends on the pressure anisotropy through the merging
velocity $V_{12}$. The integral over islands includes only
interactions for which $V_{12}$ is real ($\alpha_{12}>0$). To estimate
the scaling of $R$ we note that $N_T=\int dig_i$ is the total number
of islands in the layer, so for densely packed islands we can define a
characteristic island radius $r_N=L/2N_T$. Thus,
$R\sim\epsilon_rc_A/r_N$. Of course, $R$ can be much smaller if
$\alpha_{12}$ in the expression for $V_{12}$ approaches zero.

If $f$ were isotropic and therefore only a function of $v$, the energy
drive operator in Eq.~(\ref{fequation}) would vanish when averaged
over the angle $\zeta$. In this limit there is zero net energy gain,
consistent with Parker's equation in the incompressible limit
\citep{Parker65}. Equation (\ref{fequation}) is an equi-dimensional
equation and therefore has no characteristic velocity scale. Solutions
therefore take the form of power-laws. An important property of such
an equation is that the fluid moments of a given order completely
decouple from those of differing order and their solutions can
therefore be readily obtained from Eq.~(\ref{fequation}) in closed
form. Specifically an equation for $p_\parallel$ and $p_\perp$ can be
obtained so that $\alpha_{12}$ in the energy drive $R$ can be
evaluated explicitly. Thus, the feedback of energetic particles on the
dynamics of reconnection can be computed. In the case of no
source, sink or scattering, for example, Eq.~(\ref{fequation}) yields
$\partial p_\parallel /\partial t=2Rp_\parallel$ and $\partial p_\perp
/\partial t=-Rp_\perp$ so that $p_\parallel$ and $p_\perp$ increase
and decrease in time, respectively, but the total energetic particle
pressure $p$ increases, $\partial p/\partial t=(2R/3)(p_\parallel -p_\perp )$.

Instead of directly evaluating the full moments of
Eq.~(\ref{fequation}), we simplify the equation by ordering the
magnitudes of the rates $R$, $\nu$ and $c_A/L$. Since the scattering
represented by $\nu$ arises from the pressure anisotropy driven by
contraction, we argue that $\nu\sim R$. On the other hand as the
spectrum begins to saturate at firehose marginal stability $R$ is
reduced and $\nu$ increases. We therefore take $\nu \gg R\sim
\epsilon_rc_A/r_N \gg c_A/L$, where the latter follows because $L\gg
r_N$. The large $\nu$ assumption allows us to solve
Eq.~(\ref{fequation}) by expanding $f$ in a series of Legendre
polynomials $f=\sum_jP_j(\zeta)f_j(v)$ where $P_j$ is the $j$th order
Legendre polynomial. By the symmetry in $v_\parallel$, $f_1$ is
zero. The equation for $f_2$ follows from balancing the reconnection
drive acting on $f_0$ with the scattering operator acting on
$f_2P_2(\zeta)$, $f_2(v)=-(Rv/6\nu )\partial f_0(v)/\partial v$.  By
averaging Eq.~(\ref{fequation}) over $\zeta$, the scattering term
vanishes and the energy drive term acting on $f_2P_2(\zeta)$ yields an
equation for $f_0(v)$,
\begin{equation}
\frac{\partial f_0}{\partial t}-\frac{R^2}{30\nu}\frac{1}{v^2}\frac{\partial}{\partial v}v^4\frac{\partial}{\partial v}f_0+ \frac{c_{Aup}}{L}f_0=\dot{A}_Tf_{up}.
\label{f0equation}
\end{equation}
This equation is again of equi-dimensional form and has power-law
solutions whose individual moments can be calculated. Evaluating the
density in steady state, for example, by integrating over velocity,
the drive term vanishes and the total number of particles undergoing
acceleration $n_T$ is given by $n_T=A_Tn_{up}$, where
$A_T=\dot{A}_TL/c_{Aup}$ is the integrated area of all of the islands
in the layer. The firehose parameter needs to be self-consistently
evaluated and for this we need
\begin{equation}
p_\parallel-p_\perp=\frac{1}{A_T}\int_{-1}^1d\zeta\int_0^\infty dv\,2\pi v^2m(v_\parallel^2-\frac{1}{2}v_\perp^2)f_2(v)P_2(\zeta ).
\label{anisotropy}
\end{equation}
Using the expression for $f_2$ and noting that $v_\parallel^2-v_\perp^2/2=v^2P_2(\zeta )$, we obtain
\begin{equation}
p_\parallel-p_\perp=\frac{R}{6A_T\nu}\int_0^\infty dv\,4\pi
mv^4f_0(v)=\frac{R}{2\nu} p_0,
\label{panisotropy}
\end{equation}
where $p_0$ is the isotropic pressure calculated from $f_0$,
\begin{equation}
p_0=\frac{p_{up}}{1-R^2L/3c_{Aup}\nu}.
\label{p0}
\end{equation}
The firehose parameter becomes
\begin{equation}
\alpha\simeq 1-\frac{4\pi p_{up}}{\bar{B}^2}\frac{R/2\nu}{1-R^2L/3c_{Aup}\nu},
\label{fh}
\end{equation}
where $\bar{B}$ is the average island magnetic field strength based on
the sum in Eq.~(\ref{rdrive}). A key feature of Eq.~(\ref{fh}) is its
singular behavior when $\delta=R^2L/3c_{Aup}\nu =1$. This singularity
can be understood from the power-law solutions to $f_0$, which
describe its behavior at energies greater than that of the source
$f_{up}$. Taking $f_0\propto v^{-\gamma}$, from Eq.~(\ref{f0equation})
we obtain $\gamma (\gamma -3)=10/\delta$ so that when $\delta=1$,
$\gamma=5$. The second solution, $\gamma =-2$, corresponds to divergent
behavior and must be rejected. The singularity in Eq.~(\ref{fh})
therefore arises when $f_0\propto v^{-5}$ and corresponds to a
divergence of the pressure integral. Thus, it is clear that the
requirement that the pressure be bounded requires that $\gamma >5$ or
$\delta<1$. In deriving Eq.~(\ref{f0equation}) for $f_0$ we have
assumed large scattering so that $\nu\gg R$. On the other hand, since
island contraction drives the anisotropy, we argued previously that
$\nu\sim \epsilon_rc_A/r_N$. Thus, $\delta\sim \epsilon_rL/r_N\gg
1$. Namely, the acceleration rate should always exceed the system
convective loss rate since $L$ is much larger than the characteristic
island size. The resulting divergence of the pressure can only be
avoided if the reconnection drive $R\propto\sqrt{\alpha}$ is reduced
by its approach to firehose marginal stability, which forces
$R\ll\nu$. Unless $p_{up}$ is very large, the only way that the
firehose condition in Eq.~(\ref{fh}) can be reached is if
$\delta\simeq 1$ or $\gamma\simeq 5$ and $f_0\propto v^{-5}$.

The total energy content $W_0=3p_0/2$ of this high energy tail can be directly calculated from the pressure in Eq.~(\ref{p0}) using $\alpha \simeq 0$ and $\delta \simeq 1$,
\begin{equation}
W_0=\frac{\bar{B}^2}{4\pi}\sqrt{3\nu L/c_{Aup}}.
\label{W0}
\end{equation}
Thus, depending on the level of scattering, the total energy density
of the energetic particles is of the order of, or somewhat greater than,
the remaining magnetic energy. In a system with low initial $\beta$
equipartition between energetic particles and magnetic field is
energetically accessible. In a system with high initial $\beta$
equipartition can only be reached if the system is open such that
energetic particles can access additional sources of magnetic free
energy.

\section{Discussion}

We have derived a general equation (Eq.~(\ref{fequation})) for
particle acceleration in a bath of merging magnetic islands in a large
1-D current layer.  We demonstrated that the $E^{-1.5}$ spectrum is a
nearly universal feature of a multi-island reconnecting system for all
values of initial $\beta$ as long as the nominal acceleration time of
energetic particles is shorter than their loss rate. This is the
correct limit as long as the characteristic magnetic island radius is
much smaller than the system scale size $L$. We argue
therefore that the widely observed $E^{-1.5}$ spectrum in the
heliosphere is a natural consequence of multi-island reconnection. The
total energy content of this $E^{-1.5}$ spectrum reaches parity with
the remaining magnetic field energy in the system.

Equation (\ref{fequation}) can be readily generalized to a 2-D
system by replacing the factors $2r_i/L$ by $4r_i^2/L^2$ in the drive
term $R$. The estimate for the scaling of $R$ is unchanged. The model
loss term $c_{Aup}f/L$ should also be replaced by the convective loss
rate ${\bf u}\cdot {\bf\nabla }f$ with ${\bf u}$ the convective
velocity of the system. The arguments leading to the $f\propto v^{-5}$
also apply to the 2-D equations. In a system in which the driver $R$
is spatially non-uniform the 2-D version of Eq.~(\ref{fequation})
could then be numerically solved for the spatial distribution of
energetic particles from reconnection.  The impact of the finite
structure of magnetic islands that might develop in the third
direction remains an important open issue
\citep{Onofri06,Schreier10,Daughton11}.

\begin{figure}
\epsscale{.70}
\includegraphics[keepaspectratio,width=4.0in]{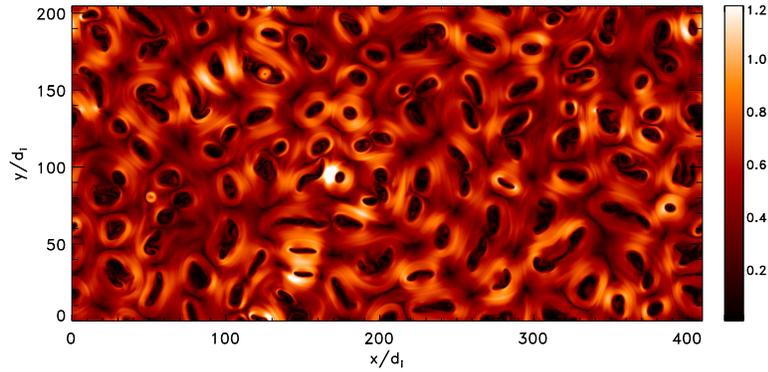}
\caption{\label{islands} The distribution of magnetic strength at late time from a multi-current layer simulation \citep{Drake10}.}
\end{figure}

\begin{figure}
\epsscale{.70}
\includegraphics[keepaspectratio,width=4.0in]{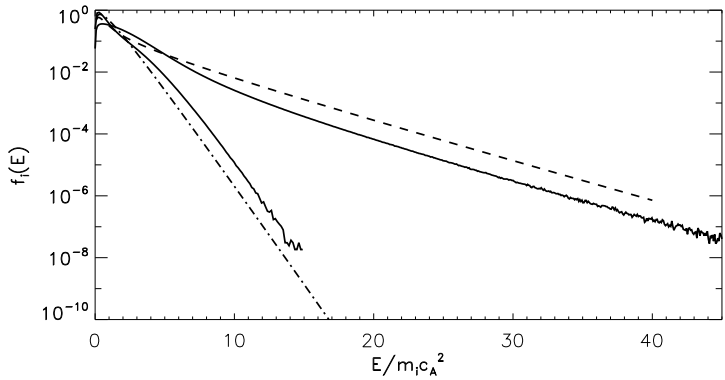}
\caption{\label{energy} The spectra of ions from the simulation of Fig.~\ref{islands} (solid lines) at $t=0, 200\Omega_{ci}^{-1}$ and from the model (dashed and dot-dashed lines).}
\end{figure}

There have now been several published simulations of particle
acceleration and associated spectra in 2-D multi-current layer systems
\citep{Drake10,Drake12}. We can compare the spectra predicted from our
equation with the results of those simulations. Since the
simulations were doubly periodic, there was no convective loss. Further,
the pressure anisotropy was strong so we consider the non-scattering
limit of Eq.~(\ref{fequation}) in which the source and loss terms are
discarded. The exact solution for $f$ is given by
\begin{equation}
f(v_\parallel^2,v_\perp^2,t)=f(v_\parallel^2e^{-2G(t)},v_\perp^2e^{G(t)},0),
\label{tequation}
 \end{equation}
where $G(t)=\int_0^td\tau R(\tau )$. This is consistent with
exponential growth of the effective parallel temperature and an
exponential decrease in the perpendicular temperature. The
omnidirectional distribution function can be computed numerically for
any specified initial distribution function for comparison with
simulation data. The comparison is made with a system with sixteen
initial current layers in a $409.6d_i\times204.8d_i$ domain, where
$d_i=c/\omega_{pi}$ is the ion inertial length \citep{Drake10}. In
Fig.~\ref{islands} we show the magnetic field strength at late time
($t=100\Omega_{ci}^{-1}$) in the simulation after islands on adjacent
current layers have overlapped. The typical island radius $r_N$ at
this time is around $15d_i$. The characteristic acceleration rate
$R\sim\epsilon_rc_A/r_N\sim 0.007\Omega_{ci}^{-1}$, where
$\Omega_{ci}$ is the ion cyclotron frequency. Reconnection remains
strong for a total time of around $100\Omega_{ci}^{-1}$ when the
pressure anisotropy shuts off reconnection. Thus, the integrated
acceleration rate is $G\sim 0.7$. The comparison between the model and the
simulation data is shown in Fig.~\ref{energy}. The particle energy
spectrum from the simulation is shown in the initial state and at
$t=200\Omega_{ci}^{-1}$ in the solid lines. Note that the initial
state is not a simple Maxwellian because of the shift in
the ion velocity distribution that is required in the current
layers. The fit of the initial spectrum with a single Maxwellian,
shown in the dot-dashed line in Fig.~\ref{energy} therefore matches
the low energy portion of the spectrum very well but underestimates
the number of particles at high energy. The late time energy spectrum
from the solution given in Eq.~(\ref{tequation}), after integration
over the angle $\zeta$ is given by the dashed line in
Fig.~\ref{energy}. The best fit corresponds to $G=0.82$ rather than
the estimate of $0.7$. The model reproduces the overall late-time
energy spectrum very well but modestly overestimates the number of
particles in the high energy tail. This is probably because the ions
in the initial spectrum have thermal speeds that are sub-Alfv\'enic so
the Fermi acceleration of the low energy ions is delayed until they gain
sufficient energy in reconnection exhausts \citep{Drake09}.

Observations in the quiet solar wind have revealed that the
super-Alfv\'enic ions display an $f(v)\propto v^{-5}$ distribution
\citep{Fisk06}. It has been suggested that solar wind turbulence would
be dissipated in reconnection current layers \citep{Servidio09} and therefore 
that reconnection is an important dissipation
mechanism in the turbulent solar wind. Solar wind observations also
reveal that the pressure anisotropy bumps against the firehose
threshold in some regions and that there are enhanced magnetic
fluctuations at these locations \citep{Bale09}. There are therefore
mechanisms in solar wind turbulence driving anisotropy and the
anisotropy is limited by enhanced scattering. Finally, the direct
observations of reconnection events in the solar wind reveal heating
but no localized regions of energetic particles
\citep{Gosling05a}. This is consistent with our picture that the
energetic particle spectrum is not produced at a single x-line but
requires that the ions interact with many reconnection sites.

The spectrum of energetic electrons in impulsive flares are not
measured {\it in situ} and must be inferred from chromospheric x-ray
emission. Nevertheless, the energetic particle fluxes do occasionally
reveal spectra as hard as $E^{-1.5}$, which corresponds to $f\propto
v^{-5}$ \citep{Holman03}. In recent over-the-limb observations of
flares in which the reconnection region high in the corona can be
directly diagnosed, it was found that all of the electrons in the
acceleration region became part of the energetic component, indicating
that all electrons in the region of energy release underwent
acceleration \citep{Krucker10}, which is consistent with our
model. The $\beta$ of these electrons was of order unity, which is
also consistent with our predictions.

Whether the sectored heliosheath magnetic field has reconnected
remains an open issue because the Voyager magnetometers are at the
limits of their resolutions at the magnetic field strengths in the
heliosheath \citep{Burlaga06a}.  Large drops in the energetic electron
and ACR population as Voyager 2 exited from the sectored zone are
consistent with reconnection as the ACR driver \citep{Opher11}. The
spectral index of the ACR particle flux measured at Voyager 1 is
slightly above $1.5$ \citep{Stone08,Decker10}. Further, the integrated
energy density of the measured ACR spectrum between $1$ and $100MeV$ and
is comparable to that of the magnetic field, which has a
magnitude of around $0.15nT$. This is again consistent with the
predictions of our model.

The equations presented here were derived in the non-relativistic
limit. However, the ideas can be easily extended to the case where the
particles are relativistic but where reconnection itself
is non-relativistic. We express the
distribution of particles in terms of the particle momentum ${\bf
  p}$. For the pressure integral to remain bounded
$\gamma >4$ for power-law distributions with
$f_0(p)\propto p^{-\gamma}$. The resulting particle flux per unit energy
interval $\Gamma$ is given by $\Gamma\propto p^2f(p)\propto
p^{2-\gamma}$. Thus, the spectrum of the flux in the strongly
relativistic limit should scale as $p^{-2}$. 

\acknowledgments

This work has been supported by NSF Grant AGS1202330 and NASA grants
APL-975268 and NNX08AV87G.

%\bibliographystyle{apj}
%\bibliography{bib}

\clearpage

\end{document}